\documentclass[twocolumn,prb,aps,showpacs]{revtex4}

\usepackage{psfrag,graphicx}
\usepackage{dcolumn}
\usepackage{amsmath,amssymb}
\usepackage{bm}
\usepackage{pstricks}
\usepackage{hyperref}
\usepackage{epsfig}

\def\nind{\noindent}
\def\<{\left<}
\def\>{\right>}
\def\ket|#1>{\left|#1\right>}
\def\bra<#1|{\left<#1\right|}
\def\elem<#1|#2|#3>{\left<#1\right|#2\left|#3\right>}
\def\({\left(}
\def\){\right)}

\def\aAg{a_{\rm Ag}}
\def\aPt{a_{\rm Pt}}

\begin{document}

\title[Short Title]{Reconstruction of the second layer of Ag on Pt(111)}

\author{Raghani Pushpa}
\affiliation{Center for Probing the Nanoscale, Stanford University, Stanford,
California 94305, USA} 
\author{Javier Rodr\'{\i}guez-Laguna}
\affiliation{Dpto. Matem\'aticas, Universidad Carlos III de Madrid,
  Madrid 28911, Spain}
\author{Silvia N. Santalla}
\affiliation{Dpto. F\'{\i}sica, Universidad Carlos III de Madrid,
  Madrid 28911, Spain}

\date{October 15, 2008}

\begin{abstract}
The reconstruction of an Ag monolayer on Ag/Pt(111) is
analysed theoretically, employing a vertically extended
Frenkel-Kontorova model whose parameters are derived from density
functional theory. Energy minimization is carried out using simulated
quantum annealing techniques. Our results are compatible with
the STM experiments, where a striped pattern is initially found which
transforms into a triangular reconstruction upon annealing. In our model
we recognize the first structure as a metastable state, while the
second one is the true energy minimum.
\end{abstract}

\pacs{
68.35.Bs, 
61.72.Bb, 
68.35.Md, 
68.55.-a, 
02.60.Pn 
}

\maketitle


\section{\label{introd}Introduction}

One of the most important physical processes that must be properly
understood in order to manufacture the nanoscale devices is surface
reconstruction, i.e., the morphology changes in the epilayer of a
material as a result of reduced coordination, either in homoepitaxial
or heteroepitaxial systems. The lattice parameter of the epilayer is
usually different from that of the bulk system, giving rise to surface
stress. This stress can be relieved in a variety of ways, inducing the
appearance of interesting patterns \cite{barthau, narasimhan, agpt}.

In this work we analyse the reconstruction of the second monolayer
(ML) of Ag deposited on a clean surface of Pt(111). The experiments of
Brune and coworkers \cite{Brune_PRB94,Brune_Nat98} show that, while
the first ML is pseudomorphic (i.e., copies the structure of Pt(111)),
the second monolayer reconstructs in a striped pattern, which upon
annealing at 800 K, gives rise to an intrincate trigonal network,
reminiscent of a kagome lattice. Along (111) direction,
silver presents a lattice parameter of $\aAg=2.95$ \AA, which is 4\%
higher than that of platinum, $\aPt=2.83$ \AA.

Henceforth, we will refer to the system composed of a complete first
monolayer (ML) of Ag on Pt(111) as Ag/Pt(111). Ag atoms in the second
ML on that structure can stay in two types of local minima, both
arranged in triangular networks, which are denoted as FCC and HCP
sites (Figure \ref{geometry}). Earlier work
\cite{Ratsch_PRB97,Fichthorn_PRL00} has shown that the two types of
sites have very similar energies, being the FCC site favoured by 3
meV. In the middle point of the segment connecting a FCC and a HCP
site there is a {\em bridge} saddle-point site. The energy of an Ag
atom in the second ML at the bridge site constitutes the energy
barrier to jump from one type of minimum to the other. It was
estimated by a theoretical fitting of experimental data on island
densities \cite{Fichthorn_PRL00} to be around 60 meV. As an important
final remark, the adatom-adatom interaction was shown to decay slowly,
with a repulsive ring past the short-range which may play an important
role in the kinetics \cite{Fichthorn_PRL00}.

In the present work we model the second monolayer of Ag on Ag/Pt(111),
referred to as Ag/Ag/Pt(111). We use a generalization of the renowned
2D Frenkel-Kontorova (FK) model \cite{Braun:book,Frenkel_39}. This
model has provided accurate and insightful analysis of similar
reconstruction processes, such as Pt(111) \cite{Raghani_PRB03} or the
first layer of Ag on Pt(111) \cite{Hamilton_PRL99}. Our model differs
from the standard FK in the following aspects. First of all, the
substrate potential is vertically extended, i.e., it becomes a
function of $x$, $y$ and $z$. This way, the film is allowed to relax
the stress by (small) vertical displacements of the atoms in the
topmost Ag layer \cite{Laguna_PRB05}. A second difference is that the
number of neighbours of any atom is not fixed beforehand, i.e., no
topology is assumed. Thus, all kinds of lattice defects are allowed in
the surface structure.

The substrate potential for the extended FK model is obtained by
choosing a physically sensible general form and obtaining its
parameters using {\em ab initio} density functional theory (DFT). For
the film potential, we use a Morse type pair potential. The obtention
of the minimum energy configurations of the extended FK model is a
complex problem, which has been approached using the recently
developed {\em replica-pinned quantum annealing} (RPQA)
\cite{Gregor_CPL05}.

The minimum energy configurations of the extended FK model are assumed
to be (local) equilibrium atomic patterns of the reconstructed
surface. Of course, this approach neglects the kinetic effects which
may be relevant in the determination of the experimental
configurations. We will provide some comments about these effects in
section \ref{discussion}.

This article is organized as follows. Section \ref{model} discusses
the DFT calculations and the obtention of the extended FK model. The
results of the numerical minimization are given in section
\ref{results}, while its physical interpretation is provided in
section \ref{discussion}, along with the most promising lines for
further work.


\section{\label{model}Model}

The behavior of an Ag monolayer on the Ag/Pt(111) system is simulated by
an extended FK model, defined by the following functional form

\begin{equation}
E(\{r_i\})=\sum_i V_s(r_i) + \sum_{\<i,j\>} V_f(|r_i-r_j|)
\label{fk.model}
\end{equation}

\nind where $r_i=(x_i,y_i,z_i)$ is the 3D position vector of the
$i^{th}$ Ag atom, $V_s$ denotes the {\em substrate potential}, $V_f$
is the {\em film potential}, and the sum over pairs $\<i,j\>$ is
extended over neighbours within a certain {\em cutoff} distance. This
cutoff distance is determined so as to include the ring of first NN
atoms. The vertical extension of the substrate potential
\cite{Laguna_PRB05} allows for stress relaxation via small vertical
displacements along $z$-direction.

\begin{figure}
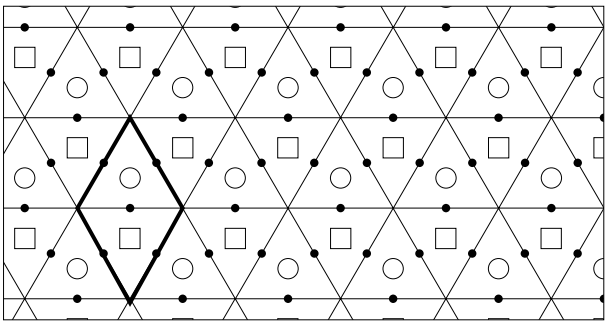

\psset{unit=1mm}
\rput(-38,-40){
\psset{unit=1.4mm,linewidth=0.1mm}
\psclip{\psset{linewidth=0mm}\psframe(-2,-2)(55,27.8)}
\rput(-5,-8.6){\pspolygon[linewidth=1.5pt](15,8.2)(10,17.2)(15,25.8)(20,17.2)
\multirput(10,0){7}{\pspolygon(0,0)(10,0)(5,8.6)
\pscircle(5,2.86){1}\psframe(9,4.73)(11,6.73)
\pscircle*(5,0){0.4}\pscircle*(2.5,4.3){0.4}\pscircle*(7.5,4.3){0.4}}}
\rput(-10,0){\multirput(10,0){7}{\pspolygon(0,0)(10,0)(5,8.6)
\pscircle(5,2.86){1}\psframe(9,4.73)(11,6.73)
\pscircle*(5,0){0.4}\pscircle*(2.5,4.3){0.4}\pscircle*(7.5,4.3){0.4}}}
\rput(-5,8.6){\multirput(10,0){7}{\pspolygon(0,0)(10,0)(5,8.6)
\pscircle(5,2.86){1}\psframe(9,4.73)(11,6.73)
\pscircle*(5,0){0.4}\pscircle*(2.5,4.3){0.4}\pscircle*(7.5,4.3){0.4}}}
\rput(-10,17.2){\multirput(10,0){7}{\pspolygon(0,0)(10,0)(5,8.6)
\pscircle(5,2.86){1}\psframe(9,4.73)(11,6.73)
\pscircle*(5,0){0.4}\pscircle*(2.5,4.3){0.4}\pscircle*(7.5,4.3){0.4}}}
\rput(-5,25.8){\multirput(10,0){7}{\pspolygon(0,0)(10,0)(5,8.6)
\pscircle(5,2.86){1}\psframe(9,4.73)(11,6.73)
\pscircle*(5,0){0.4}\pscircle*(2.5,4.3){0.4}\pscircle*(7.5,4.3){0.4}}}
\endpsclip
}
\vspace{4cm}
\caption{\label{geometry} Geometry of the Ag/Pt(111) system:
  top sites are shown as the vertices of the triangular
  network; FCC, HCP and bridge sites are shown as empty circles, empty
  squares and filled dots respectively. A unit cell is marked 
  with a thicker line.}
\end{figure}

The appropriate intensive magnitude to minimize is, of course, the
{\em energy per unit area}. The unit cell is shown by a bold line 
in Figure (\ref{geometry}). The atomic density at the
surface is, therefore, another relevant parameter to be obtained in
our method.

Let us review concisely the geometry of our model. There are four
types of distinguished points on the substrate lattice, which are
known as FCC, HCP, top and bridge, as shown in Figure
\ref{geometry}. The local equilibrium positions for an atom are the
FCC and HCP, being the first slightly favoured energetically. Bridge
sites are saddle-points (usually denoted with the letter $p$), and top
positions, so called because they stand exactly above the substrate
atoms, are strongly disfavoured energetically.

\subsection{Density Functional Theory calculations}

The parameters in the substrate potential are obtained using {\it ab
initio} density functional theory (DFT) calculations with the
pseudopotential approach and plane waves basis, as implemented in
Quantum-Espresso \cite{QE}. We use generalized gradient approximation
(GGA) for the exchange correlation interaction with the functional
proposed by Perdew, Burke and Ernzerhof \cite{PBE}. A kinetic energy
cut-off of 367.2 eV and a higher cut-off of 2940 eV is used for the
augmentation charges introduced by the ultrasoft pseudopotential
\cite{uspp}. To improve the convergence, a gaussian smearing of width
0.68 eV was adopted.  Brillouin zone integrations for the ($1\times
1$) surface cell, shown by a thick line in Figure \ref{geometry}, were
carried out using a ($15\times 15\times 1$) mesh of $k$-points.

We obtained the bulk lattice parameter for Ag and Pt as 4.17 and 4.0
\AA\ respectively, which give a misfit of 4.25\%. These values compare
well with the experimetal values, 4.09 and 3.92, with a misfit
of 4.33\%, and the previous theoretical calculations of 4.19 \AA\ and
4.01 \AA\ \cite{Ratsch_PRB97}. According to our values, the nearest
neighbour (NN) distance for Ag and Pt on the (111) surface are 2.95
and 2.83 \AA\ respectively.


\subsection{Substrate potential}

The {\em substrate potential} $V_s(r)$ must share the periodicity of
the substrate lattice, i.e., it will take the form of a Fourier series
\cite{Hamilton_PRL99},

\begin{equation}
V_{s}({\bf r},z) = V_0 + \sum_{\bf G} V_{\bf G}(z) e^{i{\bf G}\cdot{\bf r}},
\label{fourier}
\end{equation}

\nind where $\bf G$ are 2D-vectors of the reciprocal substrate lattice
and ${\bf r}=(x,y)$. This series is truncated arbitrarily to contain
the first two shells of {\bf G} vectors, of length $4\pi/\sqrt{3}\aPt$
and $4\pi/\aPt$, separated by $2\pi/3$ rad, with $\aPt=2.83$ \AA. The
dependence on $z$ is restricted to the coefficients $V_{\bf
G}(z)$. When we expand equation (\ref{fourier}) in the $V_{\bf G}$'s
using the first two shells of {\bf G} vectors and apply the symmetries
of the surface, we get an equation with four unknowns
\cite{pushpathesis}. In order to obtain these four unknowns, we need
to know the values of the substrate potential ($V_{s}$) at four
non-equivalent positions on the surface, which we choose to be the
FCC, HCP, bridge and top sites. Thus, we express the substrate
potential $V_s({\bf r})$ in terms of $V_{FCC}$, $V_{HCP}$, $V_p$,
$V_{top}$. These values are calculated within DFT for a monolayer of
Ag on a Ag/Pt(111) slab. The Ag/Pt(111) slab is simulated using a
symmetric slab of seven Pt layers and one Ag layer adsorbed on both
sides of Pt. We first calculate the equilibrium heights by relaxing
the Ag monolayer on Ag/Pt(111) along the $z$-direction at the four
sites. We use a vacuum width of 8 layers (18.47 \AA) between the two
Ag/Ag/Pt(111) slabs with the in-plane lattice parameter of 2.83 \AA.
Since the $V_{\bf G}$'s are functions of $z$, we calculate the energy
of the second ML of Ag at various values of $z$ by keeping the
monolayer at different heights from the substrate. In this way, four
functions of $z$ are obtained: ${\tilde v}_k(z)$, with $k\in\{$FCC,
HCP, p, top$\}$, by fitting the DFT results to a Morse-like form:

\begin{equation}
{\tilde v}_k(z)=
\frac{A_k}{\mu_k-\nu_k} 
\( \nu_k e^{-\mu_k(z-z_k)} -
\mu_k e^{-\nu_k(z-z_k)} \) 
+V_{k,0}
\label{morse.z}
\end{equation}

\noindent where $\nu_k=\mu_k/2$. The substrate potential is now well
defined from {\em ab initio} calculations.

If we evaluate the substrate potential at the equilibrium height of
the FCC position, we should obtain the chemical potential for the
Ag/Ag/Pt(111) system, $\mu$, i.e., the energy required to take a
single Ag adatom from its equilibrium position on the surface to
infinity. We have estimated $\mu$ from DFT calculations as three times
the binding energy of two Ag atoms in bulk. Thus, $\mu=1.251$ eV.

Table \ref{Vs:values} gives the equilibrium heights calculated from
DFT at each type of site, along with the energy required to create a
stacking fault relative to that of the FCC position. The value for
the energy difference between the FCC and the HCP sites is consistent
with the previous values given in the literature, about 3 meV. The
energy barrier between them is lower in our case, about 35 meV, as
compared to the 60 meV of other sources
\cite{Ratsch_PRB97,Fichthorn_PRL00}. The reason for this discrepancy
is that they have done calculations on a single adatom adsorbed on FCC
or bridge sites, whereas we use one full monolayer of Ag on
Ag/Pt. Using a single adatom on FCC or bridge sites would obviously
increase the binding energy of the adatom due to its lower
coordination on the surface with respect to a full monolayer. 

\begin{table}
\begin{ruledtabular}
\begin{tabular}{lllll}
                & FCC     & HCP     & bridge  & top     \cr
$z_{eq}$ (\AA)  & 2.45926 & 2.45977 & 2.54478 & 2.80834 \cr
$E_{min}$ (meV) & 0       & 3.638   & 35.2784 & 174.066 \cr
\end{tabular}
\end{ruledtabular}
\caption{\label{Vs:values}Equilibrium heights ($z_{eq}$) and corresponding
  energies to create an stacking fault in the second ML of Ag with 
respect to lowest energy FCC site, as obtained from DFT.}
\end{table}
 
\begin{table}
\begin{ruledtabular}
\begin{tabular}{lllll}
$z$ (\AA) & E (FCC) & E (HCP) & E (bridge) & E (top) \cr
\hline
2.27447 & 21.3928 & 26.7308 & 69.2716 & 326.563 \cr
2.64406 & 17.4284 & 19.2576 & 42.4184 & 192.134 \cr
2.73646 & 34.4624 & 36.7064 & 57.6368 & 178.520 \cr
3.34878 & 252.375 & 253.416 & 254.735 & 269.606 \cr
\end{tabular}
\end{ruledtabular}
\caption{\label{Vs:zdep} Energy (in meV) of an Ag atom on each of the
  four types of sites, at some values of $z$, with respect to the
  energy of the atom at its equilibrium height on FCC, as obtained
  with DFT.}
\end{table}

\begin{figure}
\epsfig{file=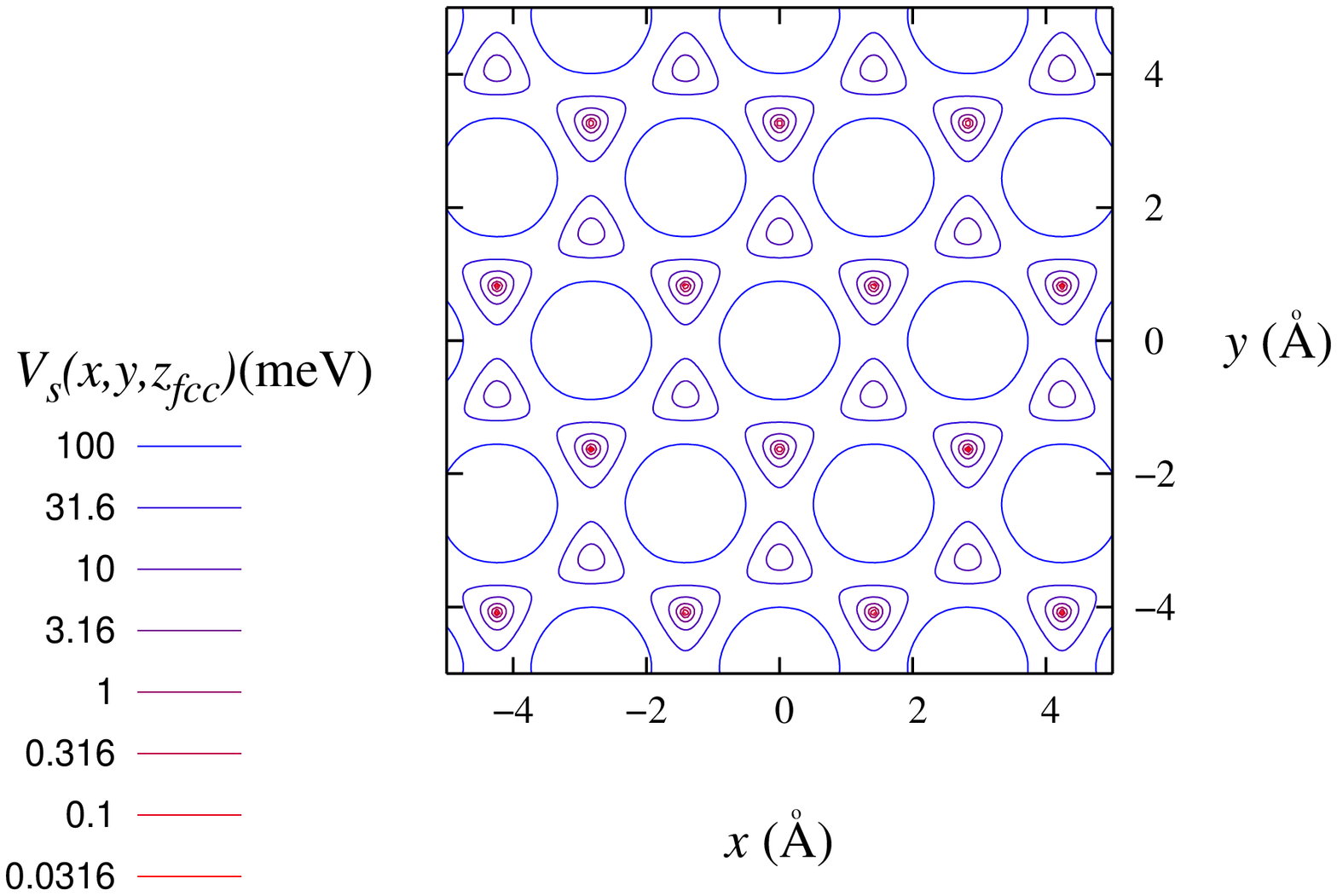,width=6cm,angle=270}
\epsfig{file=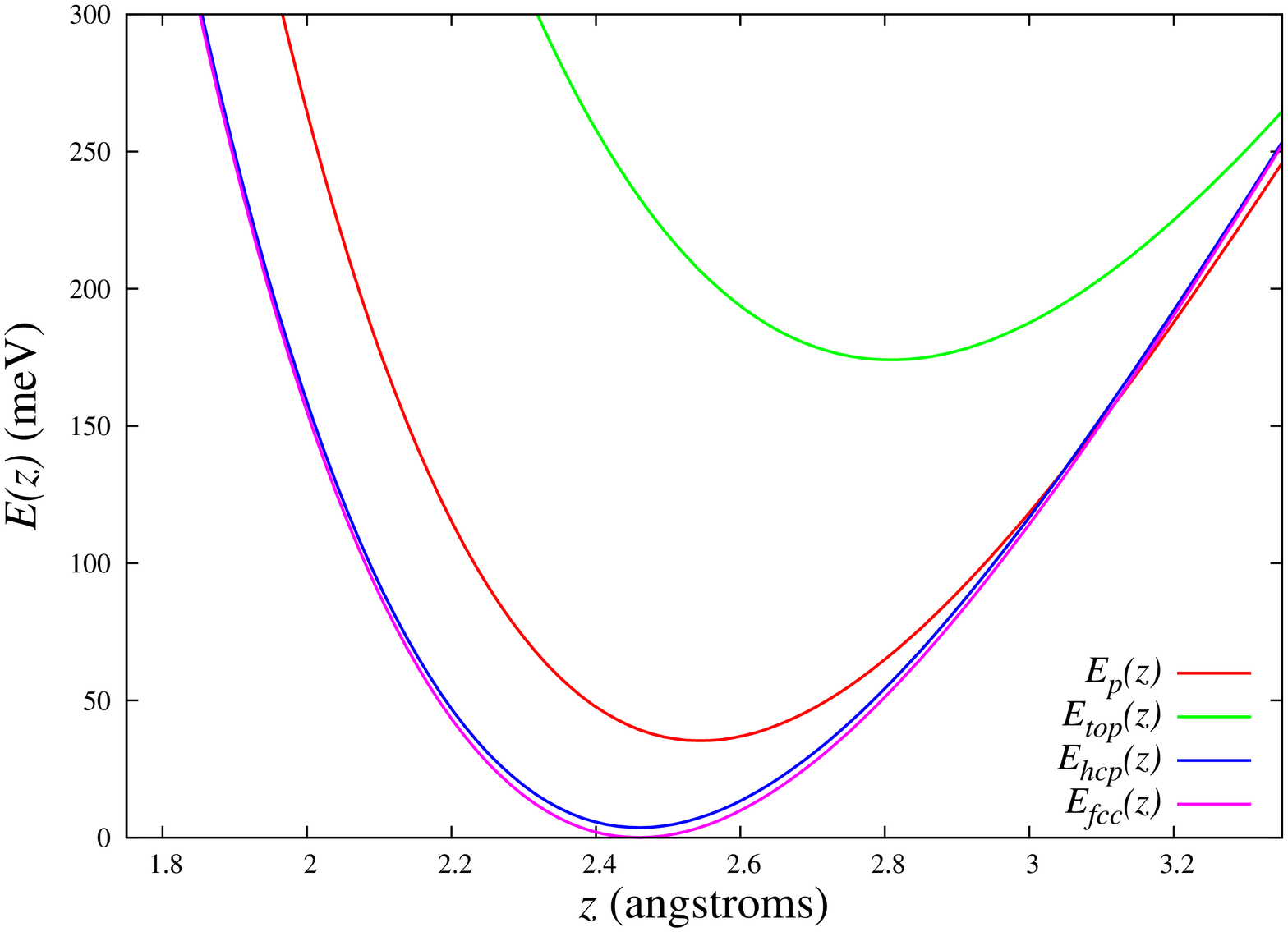,width=5.5cm,angle=270}
\caption{\label{substrate} (Color online) The substrate potential
  $V_s(r)-\mu$. Upper: contour plot of $V_s(x,y,z_{fcc})-\mu$
  calculated at the equilibrium height of the FCC sites. Lower:
  vertical dependence of the substrate potential, for the FCC, HCP,
  top and bridge sites.}
\end{figure}

Some DFT data for the dependence on $z$ of the substrate potential are
given in Table \ref{Vs:zdep}. A plot of $V_s(r)-\mu$ is shown in
Figure \ref{substrate}. We would like to remark that the fit of
$V_s(r)$ includes more points than those shown in table \ref{Vs:zdep}
(between 5 and 8 per site type), but only those are shown because they
are computed for equal values of $z$.


\subsection{Film potential}

The film potential $V_f(r)$ (also known as {\em surface potential})
simulates the interaction among Ag atoms in the Ag layer on Ag/Pt(111)
surface. We assume it to be an isotropic pair potential of Morse type.

\begin{equation}
V_f(r)=a_0\left[ \( 1-e^{-a_1(r-b)} \)^2 -1\right]  
\label{film.potential}
\end{equation}

In this equation, $b$ is the equilibrium distance, $a_0$ the depth of
the potential well and $a_1$ the curvature around the minimum.
However, $V_f(r)$ is rather difficult to obtain from DFT calculations
due to the heteroepitaxial nature of the problem. The closest
approximation would be to calculate it for Ag atoms on a Ag(111)
surface. However, the nearest neighbour (NN) distance in the Ag layer
on Ag/Pt(111) surface is 2.83\AA, while the NN atomic distance in the
Ag layer on Ag(111) surface is 2.95\AA. Therefore, as one would
expect, Ag layer on Ag/Pt(111) surface is under compressive stress of
0.198 eV/\AA$^2$ and Ag layer on Ag(111) surface is under tensile
stress of 0.091 eV/\AA$^2$. Due to this difficulty, we have fixed the
value of $b$ to 2.95\AA, i.e., the lattice parameter of Ag, and run
simulations for different values of $a_0$ and $a_1$, as we will see in
the next section.

In order to calculate the total film energy, we only consider pairs of
atoms closer than a certain cutoff distance $d_{\rm cutoff}=$6\AA. As
a technical side remark, we added a tiny constant to the potential in
order to avoid a discontinuity at that distance. Therefore, the
topology of the epilayer is not fixed beforehand: the number of
neighbours of a given atom is not restricted.

The Ag-Ag pair potential has a complex structure at longer
distances, which we have neglected in this work. Other authors
\cite{Fichthorn_PRL00} have shown the effect of Friedel-like
oscillations at long distances, recognizing a ring of repulsion in the
pair potential. This ring is beyond our cutoff distance and we will
assume it to be negligible in order to determine the equilibrium
configuration. Its effects on the kinetics of the system might be
much more pronounced, and will be discussed in section
\ref{discussion}.


\section{\label{results}Optimum configurations}

\subsection{\label{optimization}Numerical optimization}

Once the extended FK Hamiltonian has been determined, i.e., the substrate
and film potential have been found, the main task is to obtain the
positions of the Ag atoms which minimize the total energy, i.e., its
equilibrium configuration.

Global optimization of a potential energy surface (PES) with a large
amount of local minima is a highly non-trivial task. Simulated thermal
annealing is one of the most popular non-biased general-purpose
optimization methods, in which thermal fluctuations allow the system
to escape metastable states in order to achieve the real global
minimum. A recent innovation is that of {\em simulated quantum
annealing}, in which quantum fluctuations collaborate with thermal
fluctuations in this task\cite{Das_Chakrabarti:Book,Lee_JPC00}. Using
path integral concepts, the original system is replaced by a set of
{\em replicas} exploring the same PES, linked among themselves with
springs of zero natural length, at a fictitious finite
temperature. {\em Replica-pinned quantum annealing} (RPQA) is a
technical variant which has provided the best non-biased results 
to date for the determination of the global minimum energy structures
of Lenard-Jones clusters. The basic idea is to update all replicas
{\em excluding} the one which has reached the best energy so far. The
details can be found in the original article by Gregor and Car
\cite{Gregor_CPL05}.

The application of the RPQA to the obtention of the global minimum of
FK models is one of the main technical improvements developed in this
work. Both in 1D and in 2D, the efficiency of the method is higher
than the case of simulated thermal annealing. RPQA tends to explore
more thoroughly the area around the best replica so far, thus removing
defects faster than other methods. Detailed results and comparisons
will be provided elsewhere.

Of course, it is impossible in these problems to be completely sure
that the best minimum obtained is the true global minimum, but after
an exhaustive search and coincidence of diverse methods a large degree
of confidence can be assumed.


\subsection{Results}

An exhaustive search for the minimum energy configurations of the
extended FK model, as defined in equation \ref{fk.model}, has been
carried out using the procedure described in the previous section.
For all the simulations, we use periodic boundary conditions (PBC),
with different sizes for the system cell. A few important parameters
were systematically varied. On one hand, the parameters $a_0$ and
$a_1$, i.e., the depth and curvature of the film potential, for which
{\em ab initio} methods have not provided a reliable estimate. On the
other hand, the atomic density of the surface, i.e., the number of
particles per unit area.

Among the huge number of local minima, we have selected three of them
because of their special physical relevance. They are local minima of
the PES for all reasonable values of $a_0$ and $a_1$.

\begin{description}

\item{\bf(a) Unreconstructed configuration.} The homogeneous
  configuration in which all Ag atoms stay in the FCC positions on 
  Ag/Pt(111). The surface is under compressive stress. 

\item{\bf(b) Striped configuration.} In this configuration the stress
  is relaxed by placing some particles in HCP positions, as shown in
  Figure (\ref{striped:fig}) ($25\times 4$ unit cell with $24\times 4$
  Ag atoms). The increase of substrate energy is compensated by the
  decrease in the film energy due of the stretching of the horizontal
  links. The wavy pattern established has a preferred wavelength, of
  approximately $25\aPt$. Thus, there is a decrease of density as
  compared to the first Ag ML, of $4\%$.

\begin{figure}
\epsfig{file=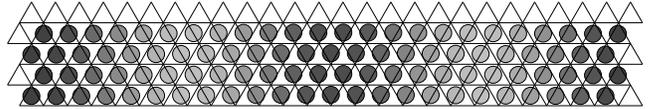,width=8cm,angle=0}
\caption{\label{striped:fig} Top view of the striped reconstruction of
the Ag/Ag/Pt(111) surface, on a $25\times 4$ unit cell. Notice the
wavy pattern, with a wavelength of $\approx 25\aPt$. Heights are
denoted by the shades of gray: the darker the atom, the lower its
height.}
\end{figure}

\item{\bf(c) Triangular configuration.} Stress is relaxed in an
  isotropic way, creating lines of defects following the symmetry axes
  of the substrate, as it is shown in Figure \ref{sun:fig}. The
  density of this configuration decreases by $7.8\%$ with respect
  to the unreconstructed case. In order to allow the relaxation of the
  surface bonds, some atoms must stay close to top sites.

\begin{figure}
\epsfig{file=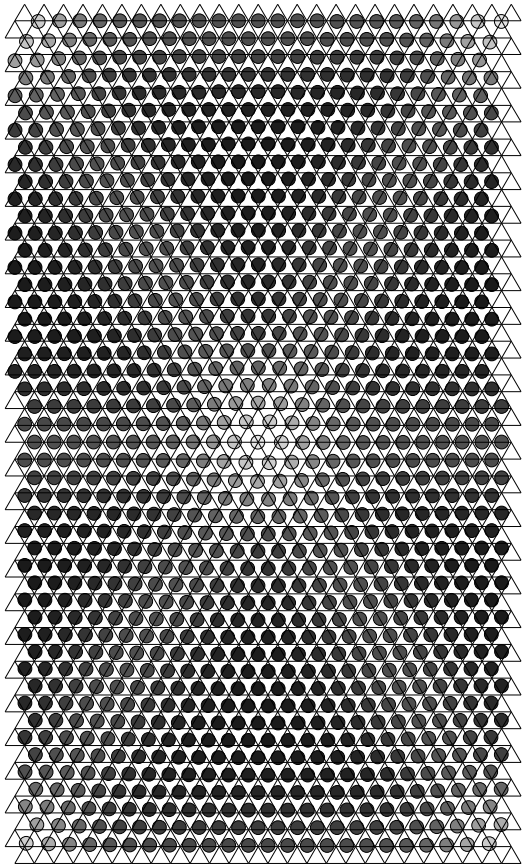,width=6.5cm,angle=0}
\caption{\label{sun:fig} Top view of the triangular reconstruction of 
Ag/Ag/Pt(111) surface. Heights are denoted by the shades of gray: the
darker the atom, the lower its height.}
\end{figure}

\end{description}

\begin{figure}
\epsfig{file=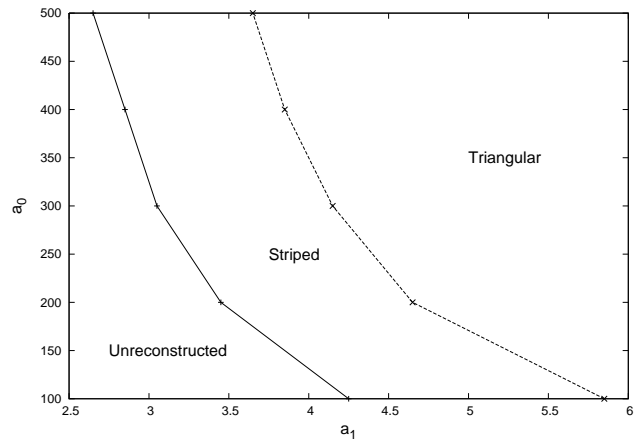,width=6cm,angle=270}
\caption{\label{phasediag}Phase diagram which shows the global minimum
of the PES as a function of the parameters of the film potential,
$a_0$ and $a_1$.}
\end{figure}

Which of them is the global minimum of the PES depends on the values
of $a_0$ and $a_1$. Figure \ref{phasediag} depicts a phase diagram in
which the stable configuration is shown as a function of these two
parameters.

The striped configuration is reported experimentally by Brune and
coworkers \cite{Brune_PRB94}. According to their work, under annealing
at 800 K, it relaxes to another configuration which is similar to our
triangular configuration. It is well known that, under annealing, the
system is much more likely to find the global energy minimum of the
PES. Therefore, the experimental data imply that the triangular
structure is, indeed, that global minimum, while the striped
configuration is a long-lived metastable state, i.e.: a local
minimum. The reasons for its long lifetime are a large energy barrier
and a small difference in energies between both configurations.

Therefore, the numerical data are compatible with the experimental
results in as much as the values of $a_0$ and $a_1$ are close to the
curve which separates the striped and the triangular regions in the
phase diagram of Figure (\ref{phasediag}). The precise point within
that curve can not be determined from our calculations.


\section{\label{discussion}Discussion and Further Work}

Our theoretical model shows from {\em ab initio} calculations, that
the Ag/Ag/Pt(111) can reconstruct into a striped or a triangular
configuration for some choices of its free parameter ($a_0$ and $a_1$
in the film potential). Both the equilibrium and the metastability
situations can be explained satisfactorily.

The fact that the striped configuration is preferred {\em before
annealing} requires further clarification. It must have a larger
energy per unit cell than the triangular structure, but kinetic
effects may make it more likely to appear. A full theoretical
explanation of this fact remains as an open task, but we may already
make some remarks with the available information. At room temperature,
Ag adatoms on the Ag/Pt(111) surface will not distinguish
clearly between HCP and FCC sites. The repulsion ring described by
Fichthorn and Scheffler \cite{Fichthorn_PRL00} explains the formation,
at low coverages, of a high amount of small islands surrounded by an
{\em exclusion area} which makes it difficult for other adatoms to
join. With the energy difference of 3 meV between FCC and HCP sites,
we can predict that, at room temperature, roughly $50\%$ of these
islands will be on each type of site. Islands will merge slowly, and
some domain boundaries will appear. At moderate coverages, the most
likely configuration for these domain boundaries is a set of parallel
lines, giving rise to the striped configuration. The complex network
which is characteristic of the triangular configuration will be more
unlikely, although it leads to a lower energy for a complete layer.

There are some differences between the global minimum with triangular
symmetry obtained in this work and the configuration reported by Brune
and coworkers. In the later structure, no atoms sit on bridge or top
sites. With our PES, the energy per unit cell of that configuration is
higher than the energy of the unreconstructed surface. When we use
this structure as the starting configuration for our quantum annealing
minimization algorithm, the system relaxes to our triangular
structure. A thorough search has been performed in the parameter space
in order to obtain the structure of Brune and coworkers as the global
minimum of the PES, without success. This could mean that, we will
have to abondon the Frenkel-Kontorova approximation in order to obtain
the experimental configuration as a global minimum of a PES.  The
assumption which is most likely to fail is the independence of the
film potential from the coordination number. The lower coordination of
some sites increases the strength with which they bind to their
neighbours. This effect is taken into account in other more complex
models, such as the glue model \cite{Ercolessi.86}.

Other systems which are of interest, where our vertically extended
Frenkel-Kontorova model can provide insight are the reconstruction of
similar metallic surfaces, such as Cu/Ru(1111). Surface growth in
semiconductors, such as the Stranski-Krastanov growth mode in 
InAs/GaAs will require major modifications, since isotropy of the film
potential might not be an acceptable approximation. On the other hand,
the vertical extension could prove much more useful in this case,
where the film tends to really curve.

On the computational side, this work applies quantum annealing methods
to a physical problem with experimental comparison, i.e., not just a
benchmark
problem\cite{Das_Chakrabarti:Book,Lee_JPC00,Gregor_CPL05}. Of course,
quantum annealing should not be considered as the panacea for all
optimization problems. It is just another tool in our numerical
toolbox, which can prove useful in a variety of cases.


\begin{acknowledgments}
The authors would like to acknowledge Shobhana Narasimhan and Stefano
de Gironcoli for very useful discussions, and also to the Scuola
Internazionale Superiore di Studi Avanzati (SISSA, Trieste, Italy),
where this work was started. This work has been partly supported by
the Spanish government through project FIS2006-04885 (J.R.-L.).
\end{acknowledgments}

\bibliography{agpt}

\end{document}